\documentclass[a4paper,fleqn,usenatbib]{mnras}

\usepackage{mathptmx}
\usepackage[T1]{fontenc}
\usepackage{ae,aecompl}

\usepackage{graphicx}	% Including figure files
\usepackage{amsmath}	% Advanced maths commands
\usepackage{amssymb}	% Extra maths symbols

\title[Supernova enrichment in low-mass clusters]{Supernova enrichment of planetary systems in low-mass star clusters}

\author[R. B. Nicholson and R. J. Parker]{
Rhana B. Nicholson\thanks{E-mail: R.B.Nicholson@2011.ljmu.ac.uk}
and Richard J. Parker
\\
Astrophysics Research Institute, Liverpool John Moores University, 146 Brownlow Hill, Liverpool, L3 5RF, UK\\
}

\date{Accepted XXX. Received YYY; in original form ZZZ}

\pubyear{2015}

\begin{document}
\label{firstpage}
\pagerange{\pageref{firstpage}--\pageref{lastpage}}
\maketitle
\begin{abstract}
The presence and abundance of short lived radioisotopes (SLRs) $^{26}$Al and $^{60}$Fe in chondritic meteorites implies that the Sun formed in the vicinity of one or more massive stars that exploded as supernovae (SNe).  Massive stars are more likely to form in massive star clusters ($>$1000 M$_{\odot}$) than lower mass clusters. However, photoevaporation of protoplanetary discs from massive stars and dynamical interactions with passing stars can inhibit planet formation in clusters with radii of $\sim$1\,pc. We investigate whether low-mass (50 -- 200 M$_{\odot}$) star clusters containing one or two massive stars are a more likely avenue for early Solar system enrichment as they are more dynamically quiescent.

%(and therefore lower density) comes after "We investigate whether low mass"!

We analyse $N$-body simulations of the evolution of these low-mass clusters and find that a similar fraction of stars experience supernova enrichment than in high mass clusters, despite their lower densities. This is due to two-body relaxation, which causes a significant expansion before the first supernova even in clusters with relatively low (100\,stars\,pc$^{-3}$) initial densities. However, because of the high number of low mass clusters containing one or two massive stars, the absolute number of enriched stars is the same, if not higher than for more populous clusters. Our results show that direct enrichment of protoplanetary discs from supernovae occurs as frequently in low mass clusters containing one or two massive stars (>20 M$_{\odot}$) as in more populous star clusters (1000\,M$_\odot$). This relaxes the constraints on the direct enrichment scenario and therefore the birth environment of the Solar System.

%The presence and abundance of short lived radioisotopes (SLRs) $^{26}$Al and $^{60}$Fe in chondrites implies the Sun formed in the vicinity of one or more massive stars that exploded as supernovae (SNe).  Massive stars are more likely to form in massive clusters (>1000 M$_{\odot}$) than lower mass clusters, however the dynamical effects from stellar interactions and photo evaporation from massive stars could be detrimental to the formation of planets. We investigate whether low mass clusters (50 - 200 M$_{\odot}$) containing one or more massive stars may be the better environment for creating Sun like stars due to being dynamically more quiet.

%We analyse the results of N body simulations that follow the evolution of low-mass star clusters containing several massive stars with a variety of initial conditions. We find that two-body relaxation occurs before the massive stars in the cluster have time to explode as SNe. On average $\sim$9 per cent of stars in the cluster were sufficiently enriched, resulting in a lower population of polluted stars from unusual low mass star clusters than high mass star clusters. Even though most enriched stars are unperturbed, the numbers that are enriched are insignificant in comparison to high mass clusters. We find that it is unlikely that these unusual low mass clusters represent the environment in which the Sun formed.

\end{abstract}


%%%%%%%%%%%%%%%%%%%%%%%%%%%%%%%%%%%%%%%%%%%%%%%%%%

%%%%%%%%%%%%%%%%% BODY OF PAPER %%%%%%%%%%%%%%%%%%

\section{Introduction}

One of the outstanding challenges in star and planet formation is to characterise the type of star formation event that formed the Sun, and hence the birth environment of the Solar System \citep[e.g.][]{2010ARA&A..48...47A,2015PhyS...90f8001P,2009ApJ...696L..13P}. Stars do not form in isolation, but rather with many other stars in regions whose densities exceed the stellar density in the Galactic field by several orders of magnitude \citep{2003ARA&A..41...57L,2010MNRAS.409L..54B}. At these high densities, dynamical interactions with other stars can truncate or destroy protoplanetary discs \citep{2001MNRAS.325..449S,2016MNRAS.457..313P}, and alter the orbits of fledgling planetary systems \citep[e.g.][]{2001MNRAS.322..859B,2002ApJ...565.1251H,2006ApJ...641..504A,2009ApJ...697..458S,2012MNRAS.419.2448P,2015MNRAS.453.2759Z}. 

Furthermore, star-forming regions often contain very massive stars, whose intense far ultra-violet (FUV) and extreme ultra-violet (EUV) radiation fields can also truncate or destroy protoplanetary discs \citep{2000A&A...362..968A,2004ApJ...611..360A}. However, in addition to their destructive properties, massive stars explode as supernovae and can chemically enrich their immediate surroundings in short-lived radioisotopes. 

Short-lived radioactive species (SLRs) with half-lives less than 10 Myr, such as $^{26}$Al and $^{60}$Fe, have been detected in calcium-aluminium-rich inclusions (CAIs) in carbonaceous chondritic meteorites in the Solar system \citep{1976GeoRL...3..109L, 2008ApJ...689..622M}. These radioactive isotopes appear to be one of the dominant sources of heat during planetesimal formation \citep{1955PNAS...41..127U, 1995Metic..30..365M} and could be important for long-term physical processes within forming planetary systems \citep{2010ARA&A..48...47A, 2016Icar..274..350L}. In comparison to the interstellar medium (ISM), the Solar system appears to contain an over abundance of these SLRs \citep{1977Icar...30..447C}. Because of the short half-lives of SLRs and their homogeneous distribution throughout the Solar System \citep{2009Sci...325..985V}, these isotopes must have been incorporated within the protoplanetary disc either shortly before or during the very early evolution of the Solar system.

SLRs can form through several mechanisms, though not all are viable methods for enriching the protosolar nebula. Cosmic ray spallation can form $^{26}$Al in large enough quantities to match the levels measured in the Solar System, but cannot produce $^{60}$Fe \citep{1998ApJ...506..898L, 2001ApJ...548.1029S}. SLRs are also produced in asymptotic giant branch (AGB) stars \citep{1994ApJ...424..412W,1999ARA&A..37..239B, 2003PASA...20..356B,  2009M&PS...44..627T}; however AGB stars are not associated with star-forming regions and a chance encounter with them is highly unlikely \citep{1994ApJ...421..605K}. The most probable scenario is that these SLRs were produced in the cores of massive young stars (>20 M$_{\odot}$), but the exact mechanism for their inclusion in the proto-Solar disc is still under debate.

One possibility is that the Sun (and its protoplanetary disc) formed from pre-enriched material \citep{2009ApJ...694L...1G, 2012ApJ...745...22G, 2012A&A...545A...4G, 2015A&A...582A..26G}. In this scenario, a first generation of stars produce SNe that pollute the surrounding medium and trigger a second generation of star formation, without destroying the giant molecular cloud (GMC). This causes the second generation of stars to form from material pre-enriched in $^{60}$Fe. Winds from a massive star in this second generation deliver $^{26}$Al into the surrounding ISM, from which a third generation of stars, including the Sun, form. However, \citet{2016MNRAS.456.1066P} show that this sequential star (and planet) formation scenario would lead to large age spreads (or even age dichotomies) due to dynamical mixing in young star-forming regions, which are not observed \citep[e.g.][]{2014prpl.conf..219S}. Furthermore, this scenario requires a substantial fraction of stars to form via triggering, which is not common in simulations of star cluster formation \citep{2015MNRAS.450.1199D}.

The other potential enrichment mechanism is direct \text{enrichment} of the protoplanetary disc from a nearby supernova explosion \citep{2000ApJ...538L.151C, 2007ApJ...662.1268O}. If enrichment occurs from a single supernova explosion, a star of 20 M$_{\odot}$ would be sufficient to deliver similar abundances to those found in the Solar System \citep{2006NuPhA.777....5W}. \cite{2000ApJ...538L.151C} and \cite{2007ApJ...662.1268O} propose that accretion of grains from the supernova ejecta could provide substantial amounts of $^{26}$Al and $^{60}$Fe, and that the high radiative luminosity could cause melting of dust grains and the formation of chondrules. The disc, when closer than 1 pc from the SN explosion, would not incur large mass loss ($<$ 1 per cent), and the disc does not absorb enough momentum from the shock to be liberated from the host protostar \citep{2007ApJ...662.1268O}. A distance to the supernova of between $\sim$0.1 -- 0.3 pc allows for sufficient enrichment without destroying too much of the disc \citep{2001MNRAS.325..449S, 2004ApJ...611..360A}, although recent work by \citet{2016arXiv160801435L} suggests that the outer distance can be relaxed somewhat.

Direct enrichment requires that the Sun formed near to high mass stars (>20 M$_{\odot}$) and due to the form of the initial mass function (IMF), it is common that several thousand low-mass stars form with every massive star. However, star clusters typically have radii of order 1\,pc \citep{2003ARA&A..41...57L}, which corresponds to initial densities of upward of 1000\,stars\,pc$^{-3}$. Stars in these dense clusters are likely to undergo significant dynamical interactions and $N$-body simulations which follow direct enrichment in massive clusters have shown that the number of G-dwarf stars that are enriched, but unperturbed by dynamical interactions is $\sim$0.5 -- 1 per cent \citep{2014MNRAS.437..946P}, making high mass clusters inefficient at enriching protoplanetary discs.

However, it is far from clear if the relationship between the most massive star in a cluster and the mass of the cluster that can form \citep{1982A&A...115...65V} is a fundamental physical outcome of the star formation process \citep{2013MNRAS.434...84W} or if it results purely from stochastic sampling of the IMF \citep{2007MNRAS.380.1271P,2008MNRAS.391..711M,2016arXiv160508438D}. The first scenario \citep{2006MNRAS.365.1333W, 2013MNRAS.434...84W} precludes the formation of massive stars (> 20 M$_{\odot}$) from forming in a low mass cluster (< 500 M$_{\odot}$). The only constraint from the latter scenario is that a star cannot form with a mass greater than that of its birth cluster \citep{2006ApJ...648..572E}. Because \citet{2013A&A...553A..31C} show that it is statistically impossible to distinguish between each scenario, we will assume that it is possible to form a massive star that explodes as a supernova in a cluster containing a few hundred low-mass stars. 

As an example, the $\gamma^2$ Velorum cluster is a low mass cluster containing 242 members \citep{2014A&A...563A..94J,2016A&A...589A..70P}, where the most massive member is $\gamma^2$ Velorum, a WC8/O8III binary with initial masses of 35 M$_{\odot}$ and 31.5 M$_{\odot}$ respectively \citep{2009MNRAS.400L..20E}. With a total cluster mass of $\sim$ 100 M$_\odot$, $ \gamma^2$ Vel provides an observational example of our simulated low-mass clusters. 

Embedded clusters typically have radii of order 1\,pc (\citealp{2003ARA&A..41...57L}, though see the more recent discussion in \citealp{2016A&A...586A..68P}), meaning that clusters with low masses will likely have a lower stellar density than their higher-mass counterparts. These lower stellar densities may imply that the stars that are enriched by supernovae suffer fewer dynamical interactions, and hence could be more viable environments for the formation and evolution of a quiescent Solar system.

In this paper we investigate whether low-mass star clusters containing massive stars can facilitate the enrichment of planetary systems without these systems suffering perturbing dynamical interactions with other stars. We determine whether the direct enrichment mechanism can occur in such low-mass clusters, and whether the fraction of enriched, unperturbed stars is higher than for the more massive clusters studied in  \citet{2014MNRAS.437..946P}. The paper is organised as follows. In Section~\ref{methods} we describe our star cluster simulations; in Section~\ref{results} we present our results; we provide a discussion in Section~\ref{discuss} and we conclude in Section~\ref{conclude}.   

\section{Methods}
\label{methods}
%We use Monte Carlo codes and N-body simulations to model and evolve clusters containing massive stars and analyse the results to see whether unusual low mass clusters are efficient producers of enriched stars.

In this section we describe our method to select low-mass clusters containing massive stars from the cluster mass function, before describing the subsequent $N$-body and stellar evolution of these clusters.   

\subsection{Creating low-mass clusters}
\begin{figure}
	\includegraphics[width=\columnwidth, scale=1.5]{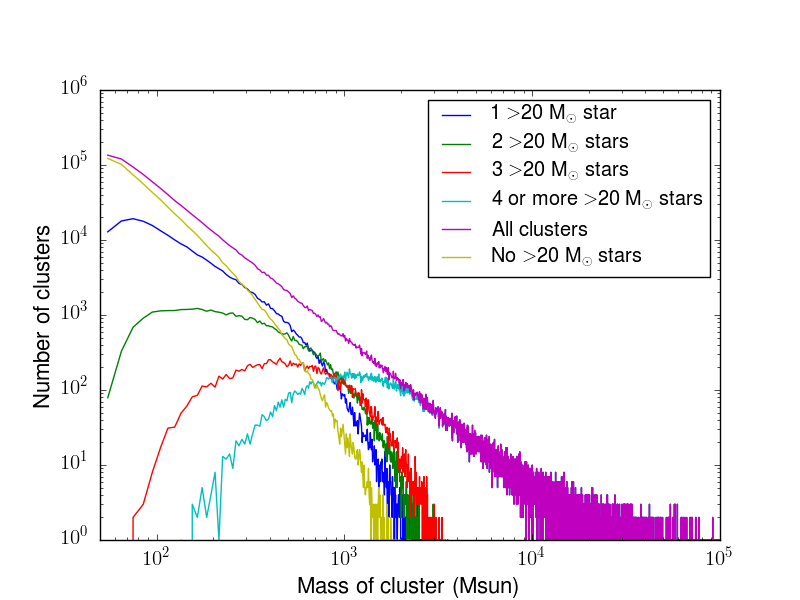}
    \caption{Star cluster mass functions for clusters containing different numbers of massive ($>$ 20 M$_\odot$) stars}
    \label{fig:clus_hist}
\end{figure}

In order to determine the fraction of low-mass clusters that contain massive stars, we randomly sample cluster masses from the following analytic fit to the observed star cluster mass function (CMF):
\begin{equation}
N(M) \propto M^{-\beta_C},
\end{equation}
where $M$ is the cluster mass and $\beta_C$ describes the slope of the observed CMF. We adopt $\beta_C = 2$ \citep{1994A&AS..104..379B,2003ARA&A..41...57L} and we sample this function for cluster masses between $50 - 10^5$\,M$_\odot$, which results in the distribution shown by the (top) magenta line in Fig.~\ref{fig:clus_hist}.

We then populate these clusters with stars drawn randomly from the initial mass function (IMF) parameterised in \citet{2013MNRAS.429.1725M}, which as a probability density function of the form: 
\begin{equation}
p(m) \propto \left(\frac{m}{\mu}\right)^{-\alpha}\left(1 + \left(\frac{m}{\mu}\right)^{1 - \alpha}\right)^{-\beta}
\label{imf}.
\end{equation}
Here, $\mu = 0.2$\,M$_\odot$ is the average stellar mass, $\alpha = 2.3$ is the \citet{1955ApJ...121..161S} power-law exponent for higher mass stars, and $\beta = 1.4$ is used to describe the slope of the IMF for low-mass objects \citep*[which also deviates from the log-normal form;][]{2010ARA&A..48..339B}. Finally, we sample from this IMF within the mass range $m_{\rm low} = 0.1$\,M$_\odot$ to $m_{\rm up} = 50$\,M$_\odot$. 

We adopt a ``soft-sampling'' technique, i.e.\,\,if the total stellar mass sampled from the IMF equals or exceeds the stipulated cluster mass, we consider the cluster to have been populated and we move on to the next cluster \citep{2006ApJ...648..572E}.

This method of randomly sampling the IMF implies that the only formal limit on the most massive star that can form is the upper limit to the IMF \citep{2007MNRAS.380.1271P}. However, several studies have proposed a fundamental relation between the cluster mass and the mass of the most massive star that can form \citep{2006MNRAS.365.1333W, 2013MNRAS.434...84W}. In the latter scenario, forming a cluster containing one massive star that can explode as a supernova requires the presence of several thousand low-mass stars ($\sim$1000\,M$_\odot$) \citep{2010ARA&A..48...47A,2014MNRAS.437..946P}.  

%\textbf{The debate whether to use random or selected sampling is an important issue with regard to the clusters selected for this paper. Sorted sampling would rule out the case where a low mass cluster could produce a massive star, due to the physical dependence on the mass of the cloud. However, as mentioned previously, observations of clusters such as $\gamma^2$ Velorum cluster reveal that cluster containing a few hundred stars can form massive stars (> 30M$_{\odot}$). Random sampling does not place limits on low mass clusters, allowing us to select the unusual outcomes of the parameter study.}

As we are interested in low-mass clusters that contain massive stars, in Fig.~\ref{fig:clus_hist} we show the CMFs for clusters that contain one star with mass $>$20\,M$_\odot$ (blue line) and two stars with masses both $>$20\,M$_\odot$ (green line). From Fig.~\ref{fig:clus_hist} it is immediately apparent that there is an equal probability of a cluster with a stellar mass of 1000\,M$_\odot$ occurring as a low-mass (50--200\,M$_\odot$) cluster containing two massive stars; low mass clusters containing one massive star are 100 times more abundant than the higher mass clusters.

Given that random sampling of the CMF and IMFs can produce low-mass clusters containing massive stars in equal or greater proportion to more massive clusters, we will explore the dynamical and evolution and enrichment probabilities in these low-mass clusters. To achieve this, we select the `median' of these low-mass clusters ($50 < M_{\rm cl}/{\rm M_\odot} < 200$), that contain the median number of stars and one or two massive ($>$20\,M$_\odot$) stars that will explode as supernovae within 10\,Myr. The median cluster containing two ($>$20\,M$_\odot$) stars contains 145 stars in total and the median cluster containing one ($>$20\,M$_\odot$) star contains 113 stars.

\subsection{$N$-body simulations}

We take the median star clusters (in terms of the number of stars) that contain one/two massive stars and evolve these in $N$-body simulations with different initial conditions. 

Both observations \citep[e.g.][]{2004MNRAS.348..589C,2009ApJ...696.2086S,2014MNRAS.439.3775G} and simulations \citep{2006A&A...449..151S,2012MNRAS.420.3264G,2013MNRAS.430..234D} of star-forming regions indicate that stars form with a hierarchical, or self-similar spatial distribution (i.e.\,\,they are substructured). It is almost impossible to create substructure through dynamical interactions; rather it is usually completely erased over a few crossing times \citep{2002MNRAS.334..156S,2014MNRAS.438..620P}.

We set up substructured star forming regions using fractal distributions, following the method of \citet{2004A&A...413..929G}. This method is described in detail in that paper, and in \citet{2010MNRAS.407.1098A} and \citet{2014MNRAS.438..620P}. 

We adopt three different fractal dimensions: $D = 1.6$ (highly substructured), $D = 2.0$ (moderately substructured) $D = 3.0$ (smooth). Note that the fractal dimensions obtained from \cite{2004A&A...413..929G} give approximate fractal values with some statistical variation around these values. Three different radii are selected for the fractals: $r_F = 0.2$\,pc, $r_F = 1.0$\,pc and $r_F = 2.0$\,pc.

%Briefly, the fractal is built by creating a cube containing `parents', which spawn a number of `children' depending on the desired fractal dimension. The amount of substructure is then set by the number of children that are allowed to mature. The lower the fractal dimension, the fewer children are allowed to mature and the cube has more substructure. Fractal dimensions in the range $D = 1.6$ (highly substructured) to $D = 3.0$ (uniform distribution) are allowed. Finally, outlying particles are removed so that the cube from which the fractal was created becomes a sphere; however, the distribution is only truly spherical if $D = 3.0$. 

The velocities of stars in the fractals are also correlated on local scales, in accordance with observations \citep{1981MNRAS.194..809L,2010A&A...518L.102A}. The children in our fractals inherit their parents' velocity, plus a small amount of noise which successively decreases further down the fractal tree. This means that two nearby stars have very similar velocities, whereas two stars which are distant can have very different velocities. Again, this is an effort to mimic the observations of star formation, which indicate that stars in filaments have very low velocity dispersions \citep{2010A&A...518L.102A}. 

Finally, we scale the velocities of the stars in the fractal to the desired virial ratio $\alpha_{\rm vir}$, where $\alpha_{\rm vir} = T/|\Omega|$; $T$ and $\Omega$ are the total kinetic energy and total potential energy of the stars, respectively. We adopt three different virial ratios in the simulations: $\alpha_{\rm vir} = 0.3$ (subvirial, or collapsing), $\alpha_{\rm vir} = 0.5$ (virial equilibrium) and $\alpha_{\rm vir} = 0.7$ (supervirial).

\begin{table*}
	\centering
	\caption{The initial conditions for all simulations where $\alpha_{\rm vir}$ is the virial ratio, D is the fractal dimension and $r_F$ is the initial radius of the cluster. All configurations of these initials conditions were run. The clusters with 113 and 145 particles contained one and two massive (>20 M$_{\odot}$) stars respectively.}
	\begin{tabular}{ccccc} % four columns, alignment for each
		\hline
		N$_{\rm stars}$ & N$_{\rm stars}$ (>20 M$_{\odot}$) & $\alpha_{\rm vir}$ & D & r$_F$\\
		\hline
		145 & 2 & 0.3, 0.5, 0.7 & 1.6, 2.0, 3.0 & 0.2, 1.0, 2.0 pc \\
        113 & 1 & 0.3, 0.5, 0.7 & 1.6, 2.0, 3.0 & 0.2, 1.0, 2.0 pc \\
		\hline
	\end{tabular}
\end{table*}

We evolve for each median cluster twenty seven sets of simulations ($r_F = 0.2, 1.0, 2.0$\,pc; $\alpha_{\rm vir} = 0.3, 0.5, 0.7$; $D = 1.6, 2.0, 3.0$) for 10\,Myr using the using the \texttt{kira} integrator within the \texttt{Starlab} environment \citep{1999A&A...348..117P, 2001MNRAS.321..199P}. In order to determine the location of the supernovae, we implement stellar evolution using the \texttt{SeBa} look-up tables \citep{1996A&A...309..179P}, which are also part of Starlab. For each realisation of the initial conditions we run 20 versions of the same simulation, identical apart from the random number seed used to set the positions and velocities, in order to gauge the inherent stochasticity in the evolution. A summary of the full parameter space of the simulations is given in Table 1.

\section{Results}
\label{results}

In this section, we will focus on one set of initial conditions for each cluster -- a subvirial ($\alpha_{\rm vir} = 0.3$), highly substructured cluster ($D = 1.6$) with initial radius $r_F = 1.0$\,pc -- before summarising the results from the other initial conditions. 

Two median clusters will be analysed, one containing a single massive star and the other containing two. We shall concentrate on the cluster containing two massive stars due to the potential for a greater enrichment fraction; however the analysis of both clusters is the same, and the median cluster containing one massive star shall be discussed in detail in Section~\ref{comparison}.

\subsection{Cluster evolution and supernova enrichment}

The median clusters chosen from our Monte Carlo sampling of the CMF and IMF resulted in a cluster containing 145 stars, of which two were massive stars (42\,M$_\odot$ and 23\,M$_\odot$). Stars with these masses (42 M$_\odot$ and 23 M$_\odot$) explode as supernovae after $\sim$4 and $\sim$8\,Myr respectively \citep{2000MNRAS.315..543H} and we therefore focus on clusters that will reach a maximum density at or around these ages (to maximise the number of low-mass stars that could be enriched by the supernovae explosions). 

Star-forming regions with medium density initial conditions ($\sim 100$\,stars\,pc$^{-3}$) and subvirial motion ($\alpha_{\rm vir} = 0.3$) will collapse to form a spherical cluster on timescales of $\sim$5\,Myr \citep{2014MNRAS.445.4037P}, and so we focus on simulations with radii $r_F = 1.0$\,pc, subvirial velocities  and a high degree of initial substructure ($D = 1.6$) in order to facilitate this cool-collapse \citep{2010MNRAS.407.1098A}.  

\begin{figure}
	\includegraphics[width=\columnwidth]{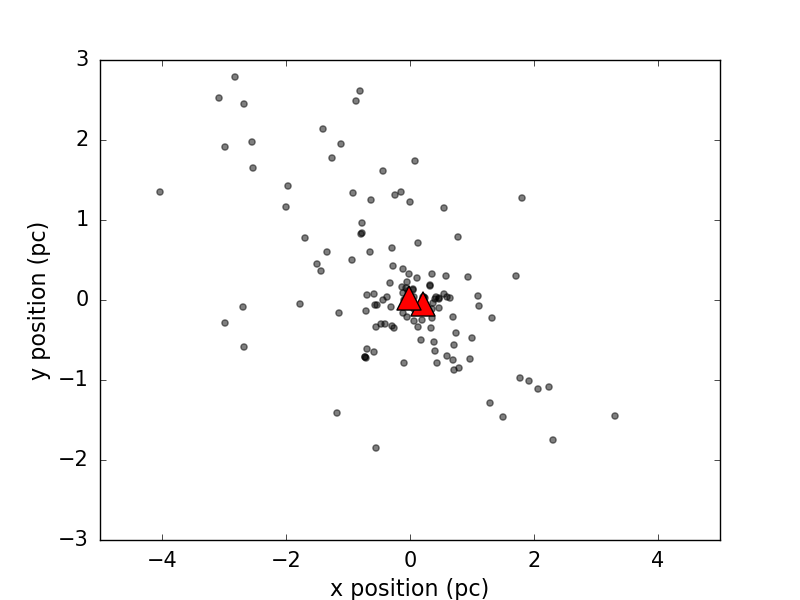}
    \caption{Snapshot of the cluster before the first SN event. The massive stars are shown by red triangles, other stars are grey points.}
    \label{fig:1st_spatial}
\end{figure}

We show an example of the morphology of a typical cluster immediately before the first supernova (at 4.4\,Myr) in Fig.~\ref{fig:1st_spatial}. The supernovae progenitors are shown by the large red triangles and have clearly migrated to the centre of the cluster. 

\begin{figure}
	\includegraphics[width=\columnwidth]{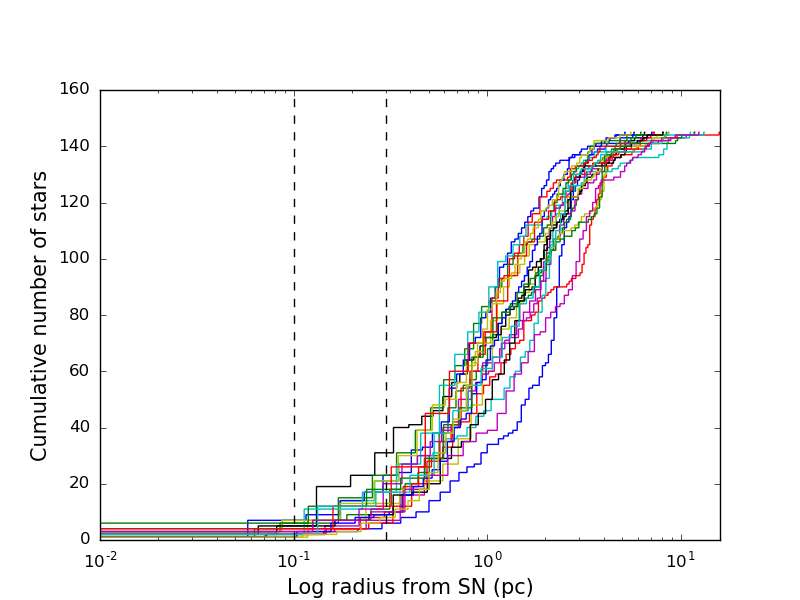}
    \caption{Cumulative distribution of stars from the first SN event. Coloured lines indicate 20 runs of the simulation. Dashed grey lines indicate 0.1 -- 0.3\,pc range where stars receive sufficient enrichment from the SN.}
    \label{fig:1st_cumul}
\end{figure}

We show the cumulative distributions of the distances from each low-mass star to the first supernova in Fig.~\ref{fig:1st_cumul}. Each of the twenty realisations of these initial conditions is shown by a coloured line, and the ``enrichment zone'' is shown by the vertical dashed lines. The enrichment zone is the distance range from the supernova (0.1 -- 0.3\,pc) thought to be most conducive to Solar system enrichment levels. Within 0.1\,pc the effects of FUV and EUV radiation are likely to destroy the protoplanetary disc, whereas beyond 0.3\,pc the amount of enrichment is too low. However, this range may be an underestimate and recent work has shown that sufficient enrichment can take place much further away from the SN \citep{2016arXiv160801435L}.

\begin{figure}
	\includegraphics[width=\columnwidth]{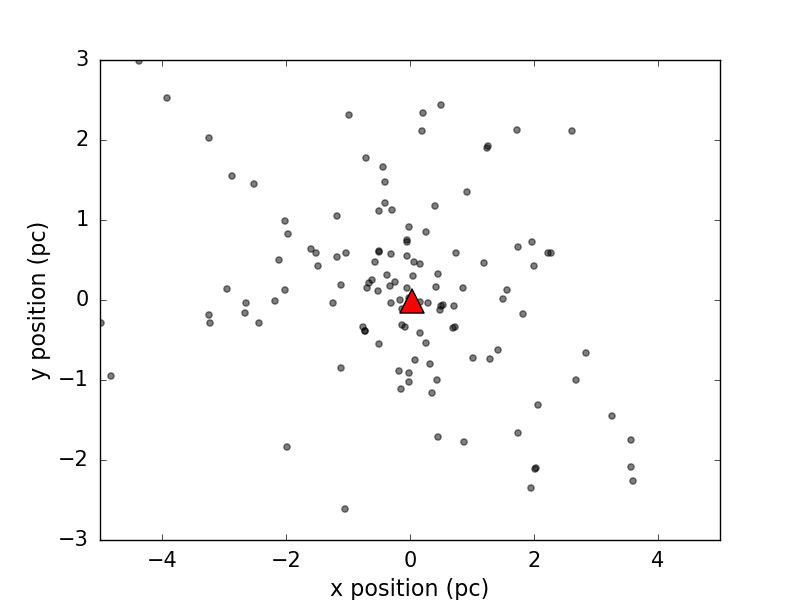}
    \caption{Snapshot of the cluster before the second SN event. The massive star is shown by a red triangle, other stars are grey points.}
    \label{fig:2nd_spatial}
\end{figure}

The average number of stars that lie within the enrichment zone during the first supernova is 9 out of 145 stars. We now consider the enrichment probability from the second, later supernova which explodes at 7.8\,Myr. The spatial distribution of the cluster at this time is shown in Fig.~\ref{fig:2nd_spatial}. It is already apparent from this figure that the cluster has expanded somewhat in the $\sim$3.4\,Myr between supernovae explosions. Again, the supernova progenitor is indicated by a red triangle in Fig.~\ref{fig:2nd_spatial}.

\begin{figure}
	\includegraphics[width=\columnwidth]{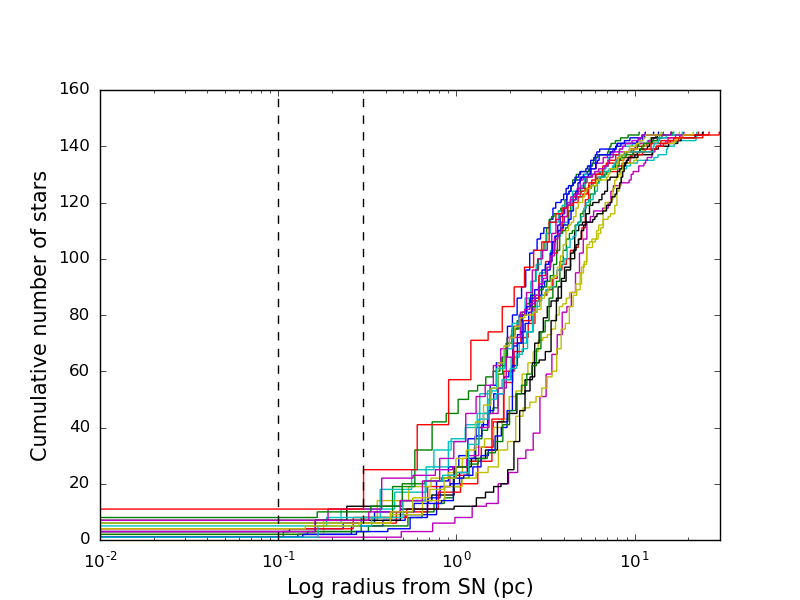}
    \caption{Cumulative distribution of stars from the second SN event. Coloured lines indicate 20 runs of the simulation. Dashed grey lines indicate 0.1 - 0.3\,pc range where stars receive sufficient enrichment from the SN.}
    \label{fig:2nd_cumul}
\end{figure}

The cumulative distribution of distances from the second supernova is shown in Fig.~\ref{fig:2nd_cumul}. This second supernova event at 7.8 Myr polluted an average of 4 stars and from inspection of Figs.~\ref{fig:1st_cumul}~and~\ref{fig:2nd_cumul} we clearly see that the fraction of polluted stars from the first supernova is more than double the fraction polluted by the second event. We also determine the number of stars polluted by both supernovae and find this to be negligible.

\subsection{Dynamical histories of polluted stars}

The bare fraction of stars that are enriched by supernovae pollution is insufficient to determine whether these systems could nurture a young Solar system. Dynamical encounters in natal star-forming environments occur frequently, and can disrupt protoplanetary discs and/or dramatically alter the architecture of fledgling planetary systems.

\begin{figure}
	\includegraphics[width=\columnwidth]{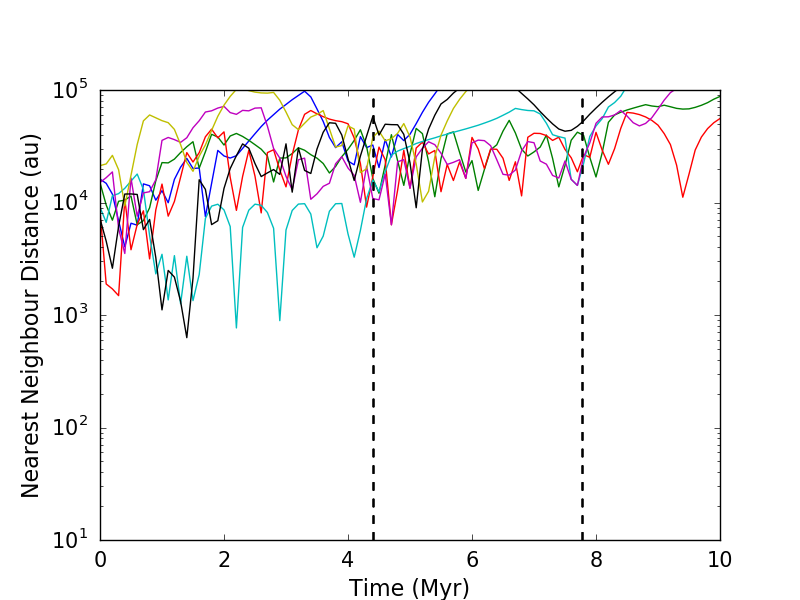}
    \caption{The closest interaction history for every enriched star with time. The black dashed lines show the times of the SN events.}
    \label{fig:interaction}
\end{figure}

For this reason, we track the dynamical history of each star that is enriched by the supernova and in Fig.~\ref{fig:interaction} we plot the nearest neighbour distance from each enriched star at all times in the simulation. Each coloured line represents every enriched star in one of our default $N = 145$, $\alpha_{\rm vir} = 0.3$, $D = 1.6$, $r_F = 1.0$\,pc simulations. The vertical dashed lines indicate the time of the two supernovae.

Two stars (the cyan and black lines in Fig.~\ref{fig:interaction}) undergo interactions that could potentially disrupt the planetary system, where the perturbing star encroaches within 1000\,AU. However, in contrast to the simulations of very dense clusters in \citet{2014MNRAS.437..946P} where 80\,per cent of enriched stars are affected by previous/susbequent interactions, only a small fraction of enriched stars (15\,per cent) in our lower density initial conditions suffer perturbing interactions.

\subsection{Other initial conditions}

%Despite the low? number of enriched stars that are perturbed by prior or subsequent dynamical interactions, the total fraction of stars that are enriched by the supernova(e) is 

In our default model ($\alpha_{\rm vir} = 0.3$, $r_F = 1.0$\,pc, $D = 1.6$) we find that the median total number of stars that are enriched per event is $\sim$13, which is comparable to the maximum number of \emph{G-dwarfs} (stars with similar mass to the Sun, 0.8 -- 1.2\,M$_\odot$) that were enriched in the more populous clusters studied in \citet{2014MNRAS.437..946P} as is the fraction of enriched G-dwarf stars that were unperturbed ($\sim$ 0.5 per cent).

In an attempt to increase the fraction of enriched stars, we explore a host of initial conditions for the star clusters. However, we find that no initial conditions result in a significantly higher (> twice) fraction of enriched stars. The principal reason for this is the effects of two-body relaxation. In order to enrich more than half of the stars in a low-mass cluster, the required stellar density \emph{at the time of the supernova(e) explosion(s)} corresponds to a radius of only $\sim$ 0.4pc, which is impossible to achieve after 5 Myr of evolution.

\begin{figure}
	\includegraphics[width=\columnwidth]{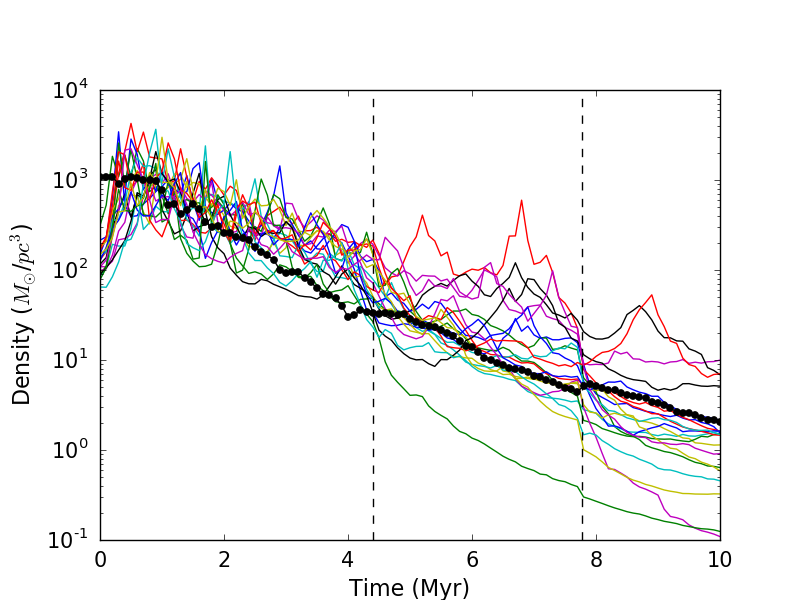}
    \caption{Evolution of the median local volume density in each simulation as a function of time. Coloured lines indicate each of the 20 simulations and the black line indicates the average central density. Vertical dashed grey lines indicate times of the supernovae.}
    \label{fig:density}
\end{figure}

Two-body relaxation results from a stellar system attempting to reach dynamical equilibrium and even a moderately dense system (100\,stars\,pc$^{-3}$) such as our default model expands significantly in the first 10\,Myr, as shown in Fig.~\ref{fig:density}. Here, the coloured lines represent the median local stellar density with time in each of the 20 realisations of these initial conditions. The solid black line is the mean central density across all 20 simulations (where we define the central density as the density within the half-mass radius). Clearly, the clusters have expanded significantly before the supernovae explosions (shown by the vertical dashed lines). 

Irrespective of the initial conditions, the clusters never attain significant enrichment because the initial density is either too high (and two-body relaxation causes significant expansion), or the initial density is so low that despite the initial subvirial motion of the stars, significant clustering of the stars never occurs. The fraction of enriched unperturbed G-dwarf stars across all initial conditions ranges between $\sim$ 0.07-0.55 per cent.

\subsection{Low mass clusters containing one massive star}
\label{comparison}

The second median cluster chosen from our Monte Carlo sampling contained 113 stars, of which one was massive (24\,M$_\odot$), resulting in a supernova explosion at $\sim$ 7 Myr. The same set of initial conditions are focused on ($\alpha_{\rm vir} = 0.3$, $r_F = 1.0$\,pc, $D = 1.6$) before summarising the others. 

An average of six stars were enriched by the supernova, a similar number to the median cluster containing two massive stars. The dynamical interaction and cluster density histories were also similar, resulting in $\sim$ 0.1-0.4 per cent of G-dwarfs being enriched and unperturbed. With almost identical enrichment fractions and dynamical histories, the difference between clusters containing one or two massive stars is small.

Fig.~\ref{fig:clus_hist} indicates that there are $\sim$ ten times as many clusters containing one massive star than clusters containing two. Intermediate mass star clusters ($\sim$ 1000 stars), such as the simulated clusters in \cite{2014MNRAS.437..946P}, are rarer, with nearly 100 low mass clusters containing one massive star for every $\sim$ 1000 M$_{\odot}$ cluster. When integrating the numbers of enriched stars produced by low mass star clusters containing one or two massive star(s) to the massive clusters that \cite{2014MNRAS.437..946P} simulated, the number of enriched stars is of the same order.

\section{Discussion}
\label{discuss}

The results of our parameter study suggest that low-mass clusters ($M_ {\rm clus} <$200\,M$_\odot$) containing one or two massive stars occur much more frequently than intermediate mass clusters ($M_{\rm clus} \sim 1000$\,M$_\odot$), which have previously been thought to be more conducive to isotopic enrichment \citep{2010ARA&A..48...47A,2014MNRAS.437..946P}. When the clusters are evolved as $N$-body simulations, we find that the fraction of enriched Sun-like stars (0.8 -- 1.2\,M$_\odot$) is lower ($\sim$ 0.1-0.6 per cent) in contrast to the simulations in \citet{2014MNRAS.437..946P} that contain $\sim$2100 stars and enrich around 1\,per cent of Sun-like stars. However, because the ratio of low mass clusters containing one or two massive stars to intermediate mass clusters is $\sim$ 1:100, the raw numbers of enriched stars are comparable.

Our choice of default initial conditions for the $N$-body simulations ($r_F = 1.0$\,pc, $D = 1.6$, $\alpha_{\rm vir} = 0.3$) results in an initial density of 100\,stars\,pc$^{-3}$. With these densities, we expect a cluster to form on timescales similar to the first supernovae in the median two massive star cluster \citep[4 -- 5\,Myr,][]{2014MNRAS.445.4037P, 2014MNRAS.438..620P} and hence expect enrichment for a high fraction of stars. However, two-body relaxation completely dominates the evolution of these low-mass star-forming regions and the median stellar density decreases so that the fraction of enriched stars is similar to that for the higher-mass clusters. 

Therefore, the fraction of stars enriched in low-mass clusters is not greater than in high mass clusters. However, due to the high number of low mass clusters containing one or two massive stars, they are as viable environments for the initial conditions for Solar System formation as high-mass clusters. Dynamical interactions between passing stars that could disrupt or destroy the natal Solar System also occur less frequently in lower-mass clusters, and the effects of photoevaporation and truncation from other massive stars are also reduced, potentially making low mass clusters a more viable birth environment for the Solar System. Our results are in broad agreement with the conclusions in \cite{2009ApJ...696L..13P}, who suggest that the Sun's birth cluster had a likely mass of between 500 - 3000 M$_{\odot}$ . We suggest that this constraint can be relaxed even further, and that the Sun could have originated in a cluster with mass between 100 - 3000 M$_{\odot}$.

Whilst most star clusters are known to disperse over time, several authors have suggested the relatively high mass M67 open cluster \citep{2005MNRAS.363..293H} as being a likely birthplace of the Sun, based on chemical composition (\citealp{2011A&A...528A..85O}, although see \citealp{2016arXiv160803788L}) and dynamical constraints \citep{2016arXiv160502965G}. Further studies that can constrain the dynamical history of M67 would be beneficial in assessing the likelihood of the Sun originating in this cluster. However, we have shown that the Sun is equally as likely to obtain the required isotopic enrichment levels in a low mass cluster (50 - 200 M$_{\odot}$) than in (less common) high mass clusters. Present and future facilities, such as the Gaia mission, may eventually be able to constrain the birth environment of the Sun by tracing it and other stars with similar chemical properties back to their formation site \citep{2009ApJ...696L..13P, 2016MNRAS.457.1062M}).  For the moment, we appeal to observational examples of low-mass clusters (50 -200 M$_{\odot}$), such as $\gamma$ Velorum, which we suggest could be equally plausible birth environments for the Solar System compared to higher-mass star clusters.

\section{Conclusions}
\label{conclude}

We analyse Monte Carlo and N-body simulations of star cluster formation and evolution to determine whether enrichment of protoplanetary discs with short-lived radioisotopes can occur from core-collapse supernovae in low-mass star clusters. Our conclusions are the following:

(i) Populating star clusters drawn from the observed cluster mass function with stars drawn randomly from the stellar initial mass function leads to a significant fraction of low-mass ($<$200\,M$_\odot$) clusters that contain one or two massive stars. These low-mass clusters with massive stars occur far more often than 1000\,M$_\odot$ clusters, which have previously been thought to be the most likely environments for enrichment from supernovae to occur.

(ii) We evolve a typical low-mass cluster using $N$-body models with stellar evolution for a wide range of initial conditions (initial radius, virial ratio, degree of substructure) and find that the typical fraction of unperturbed enriched G-dwarf stars is of order $\sim$ 0.1-0.6 per cent, similar to the fraction of unperturbed enriched G-dwarfs in high-mass clusters.

(iii) The principal reason for the low fraction of enriched stars in both low and high-mass clusters is two-body relaxation. Even if the supernova \text{enrichment} occurs at early times ($\sim$4\,Myr), the cluster has already expanded and the stellar density has decreased by a factor of ten. Stellar densities that do not result in significant expansion from two-body relaxation are inherently so diffuse that little or no enrichment can occur.

(iv) As a percentage, the fraction of enriched stars in low-mass clusters is similar to the fraction of enriched stars in more populous clusters. However, the higher number of low mass clusters from the cluster mass function means that the raw number of enriched stars that enter the Galactic field from lower-mass clusters is similar to intermediate mass clusters.

Based on these conclusions, we reduce the constraints placed on the direct enrichment scenario and find that it is as effective in low mass clusters ($M_{\rm clus} \sim 50-200$\,M$_\odot$) as in populous clusters ($M_{\rm clus} \sim 1000$\,M$_\odot$). It is likely that its effectiveness will decrease in more massive clusters where disc truncation or destruction from EUV and FUV radiation from massive stars will dominate. We will explore the implications of this in a future paper.  

\section*{Acknowledgements}

We thank the referee Simon Portegies Zwart for a positive and constructive report. We thank Tim Lichtenberg for helpful comments and suggestions and Sam Walton for coding advice. RBN is partially supported by an STFC studentship. RJP acknowledges support from the Royal Astronomical Society in the form of a research fellowship.

%%%%%%%%%%%%%%%%%%%%%%%%%%%%%%%%%%%%%%%%%%%%%%%%%%



\bibliographystyle{mnras}
\bibliography{bib} % if your bibtex file is called example.bib

%%%%%%%%%%%%%%%%%%%%%%%%%%%%%%%%%%%%%%%%%%%%%%%%%%

%%%%%%%%%%%%%%%%%%%%%%%%%%%%%%%%%%%%%%%%%%%%%%%%%%

% Don't change these lines
\bsp	% typesetting comment
\label{lastpage}
\end{document}